\documentclass[aps,prb,twocolumn,superscriptaddress,floatfix]{revtex4}

\usepackage{graphicx}
\usepackage{dcolumn}
\usepackage{bm}

\begin{document}

\title{Fermi Velocity Enhancement in Monolayer and Bilayer Graphene}
\author{Giovanni Borghi}
\affiliation{International School for Advanced Studies (SISSA), via Beirut 2-4, I-34014 Trieste, Italy}
\author{Marco Polini}
\email{m.polini@sns.it}
\affiliation{NEST-CNR-INFM and Scuola Normale Superiore, I-56126 Pisa, Italy}
\author{Reza Asgari}
\affiliation{School of Physics, Institute for research in fundamental sciences, IPM 19395-5531 Tehran, Iran}
\author{A.H. MacDonald}
\affiliation{Department of Physics, University of Texas at Austin, Austin TX 78703, USA} 

\begin{abstract}
In single-layer graphene sheets non-local interband exchange leads to a renormalized Fermi-surface 
effective mass which vanishes in the low carrier-density limit.  We report on a comparative study
of Fermi surface effective mass renormalization in single-layer and AB-stacked bilayer graphene.  
We explain why the mass does not approach zero in the bilayer case, although its value is still strongly suppressed.   
\end{abstract}

\maketitle

\section{Introduction}  
Graphene is an atomically thin two-dimensional electron system composed of carbon atoms 
on a honeycomb lattice.  Experimenters have recently~\cite{reviews} made progress with techniques which
enable isolation and 
study of systems with one or a small number of graphene layers.  
Much of the fundamental physics interest in 
graphene systems follows from the fact that its envelope-function low-energy  
Schr\"odinger equation is equivalent to the massless limit of a two-dimensional 
Dirac equation.  In the case of graphene the spinor structure in the Dirac equation
refers to honeycomb-sublattice and Brillouin-zone valley, instead of spin and 
electron-positron, degrees of freedom.  Graphene therefore presents a new type of 
many-body problem in which single-particle quantum mechanics is relativisitc, but 
interactions are essentially non-relativistic instantaneous Coulomb interactions.
 
One of the key consequences of electron-electron interactions in single-layer 
graphene is an enhancement of the quasiparticle Fermi velocity~\cite{velocityrenorm,chirality_correlations}
which diverges logarithmically as the carrier density goes to zero.  For moderate interaction strengths~\cite{ourrpa,dassarmadgastheory} the velocity 
enhancement (Fermi liquid effective mass suppression) is dominated by non-local interband exchange effects 
which are most easily explained using the pseudospin language~\cite{chirality_correlations,min_prb_2008} explained below. 
In this article we present a close comparison between the physics of mass renormalization in single-layer 
and AB-stacked bilayer graphene, explaining why the velocity renormalization is finite in the low-density limit in the latter case.     

\section{Pseudospin Description of Velocity Renormalization in Single-Layer Graphene}

We begin by recalling the continuum-model Hamiltonian ($\hbar=1$) of massless Dirac fermions interacting with non-relativistic Coulomb interactions.
The Hamiltonian has one-particle band-energy and two-particle Coulomb-interaction contributions:  
\begin{eqnarray}\label{eq:hamiltonian}
{\hat {\cal H}} &=& v \sum_{{\bm k}, \alpha, \beta} {\hat \psi}^\dagger_{{\bm k}, \alpha} ({\bm \sigma}_{\alpha\beta} \cdot {\bm k}) {\hat \psi}_{{\bm k}, \beta}\nonumber\\
&+&\frac{1}{2S}\sum_{{\bm q} \neq 0} \sum_{{\bm k}, {\bm k}'}\sum_{\alpha,\beta} v_q 
{\hat \psi}^\dagger_{{\bm k}-{\bm q}, \alpha} {\hat \psi}^\dagger_{{\bm k}'+{\bm q}, \beta}
{\hat \psi}_{{\bm k}', \beta} {\hat \psi}_{{\bm k}, \alpha}~,
\end{eqnarray}
where $v$ is the bare Fermi velocity, $\alpha, \beta$ are labels for the two honeycomb sublattices which we treat as the labels of a quantum 
spin-$1/2$-like (pseudospin) degree-of-freedom, $S$ is the area of the system, and $v_q$ is the 2D Fourier transform of the interparticle potential.  
(Because our analysis of Fermi velocity 
renormalization is based on a screened exchange approximation, in which different spins and valleys are 
independent, we do not explicitly account for these degrees of freedom.)  From 
Eq.~(\ref{eq:hamiltonian}) we see that in pseudospin language the
band Hamiltonian consists of a momentum-dependent pseudospin effective magnetic field
which acts in the direction of momentum ${\bm k}$.  The band eigenstates in the positive and 
negative energy bands have their pseudospins either aligned or opposed to the direction of 
momentum. 

When interactions are treated in a mean-field approximation we obtain~\cite{Giuliani_and_Vignale} 
\begin{equation}\label{eq:hf_interaction}
{\hat {\cal H}}_{\rm MF} = -\frac{1}{S} \sum_{{\bm k}, {\bm k}'} \sum_{\alpha, \beta} v_{{\bm k}-{\bm k}'}
\; \rho_{\alpha\beta}({\bm k}') \;  
{\hat \psi}^\dagger_{{\bm k}, \alpha}{\hat \psi}_{{\bm k}, \beta}~.
\end{equation}
where the density matrix $\rho_{\alpha\beta}({\bm k}') = \langle \Psi_0 |{\hat \psi}^\dagger_{{\bm k}, \beta}{\hat \psi}_{{\bm k}, \alpha}|\Psi_0 \rangle$ and
$|\Psi_0\rangle$ is the mean-field-theory ground state.  
(In both single-layer and bilayer cases we assume that translational symmetry is not broken so that the electron density is constant and the 
Hartree contribution to the mean-field Hamiltonian can be ignored.) 
We parametrize the pseudospin-density matrix $\rho_{\alpha\beta}({\bm k})$ in terms of charge and pseudospin-density contributions:
\begin{eqnarray}\label{eq:pseudospin-density-matrix}
\rho_{\alpha\beta}({\bm k}) &=& \frac{n^{(0)}_{{\bm k}, +} + n^{(0)}_{{\bm k}, -}}{2} \; \delta_{\alpha\beta} \nonumber \\ 
&+& \frac{n^{(0)}_{{\bm k}, +} - n^{(0)}_{{\bm k}, -}}{2} \; {\hat {\bm n}}({\bm k})\cdot {\bm \sigma}_{\beta\alpha}
\end{eqnarray}
where $n^{(0)}_{{\bm k}, s}$ are noninteracting band occupation factors.  Assumming that the valence band (label $s=-1$) is 
full and that spatial isotropy is not broken, the occupation factors are completely fixed by
the carrier density $n$. For example, for a n-doped system at $T=0$ $n^{(0)}_{{\bm k}, s} = \Theta(k_{\rm F}- k)$ for $s =+1$ and $n^{(0)}_{{\bm k}, s} =1$ for $s =-1$, where $k_{\rm F} \propto \sqrt{n}$ is the Fermi wavevector and $\Theta(x)$ is a step function. 
The unit vector ${\hat {\bm n}}({\bm k})$ specifies the pseudospin orientation 
of the positive energy band at momentum ${\bm k}$.

Using Eq.~(\ref{eq:pseudospin-density-matrix}) in Eq.~(\ref{eq:hf_interaction}) and adding the band kinetic energy 
one can easily find that the total mean-field Hamiltonian is 
\begin{equation}\label{HF_hamiltonian}
{\hat {\cal H}}_{\rm HF} = \sum_{{\bm k}, \alpha, \beta} {\hat \psi}^\dagger_{{\bm k}, \alpha} 
[\delta_{\alpha\beta}B_0({\bm k}) +{\bm \sigma}_{\alpha\beta} \cdot {\bm B}({\bm k}) ]{\hat \psi}_{{\bm k}, \beta},
\end{equation}
where the self-consistent Hartree-Fock fields are defined by
\begin{equation}\label{eq:bnot}
B_0({\bm k}) = - \int \frac{d^2{\bm k}'}{(2\pi)^2} v_{{\bm k}-{\bm k}'} \; [f_+(k') -1/2 ]~,
\end{equation}
and
\begin{eqnarray}\label{eq:hf-b-field}
{\bm B}({\bm k}) =  v{\bm k}- 
\int \frac{d^2{\bm k}'}{(2\pi)^2} v_{{\bm k}-{\bm k}'} f_-(k') {\hat {\bm n}}({\bm k}')~.
\end{eqnarray}
In Eq.~(\ref{eq:bnot}) we have subtracted the neutral system value of $B_0$, which is easily seen
to be wavevector independent.  This subtraction 
is necessary in order to make $B_0$ finite and represents only a convenient choice for the zero of energy. 
In Eq.~(\ref{eq:hf-b-field}) the first term 
is the band-structure pseudospin magnetic field, and 
the second term is the exchange field. We also have introduced the short-hand notation
\begin{equation}
f_\pm(k) = \frac{n^{(0)}_{{\bm k}, +} \pm  n^{(0)}_{{\bm k}, -}}{2}~,
\end{equation}
which recognizes that $n^{(0)}_{{\bm k}, s}$ depends only on $k=|{\bm k}|$.

We now demonstrate that the solution of the Hartree-Fock equation is
\begin{equation}\label{eq:equilibrium_B}
{\bm B}({\bm k})= v{\bm k} - \int \frac{d^2{\bm k}'}{(2\pi)^2} v_{{\bm k}-{\bm k}'} f_-(k') {\hat {\bm k}}'~.
\end{equation}
This result is a consequence of the isotropic interaction $v_{{\bm k}-{\bm k}'}=v(|{\bm k}-{\bm k}'|)$ 
which can be decomposed in angular momentum components using 
\begin{equation}\label{eq:am_expansion}
v_{{\bm k}-{\bm k}'} = \sum_{m=-\infty}^{+\infty}~V_m(k,k')e^{i m (\varphi_{\bm k}-\varphi_{{\bm k}'})}
\end{equation}
with
\begin{eqnarray}
V_m(k,k') &=& \int_0^{2\pi} \frac{d\theta}{2\pi} e^{-i m \theta} \nonumber \\
&\times &v(\sqrt{k^2+k'^2-2kk'\cos(\theta)})~.
\end{eqnarray}
Substituting this representation in Eq.~(\ref{eq:equilibrium_B}), 
performing the angular integration over $\varphi_{{\bm k}'}$
we find that 
\begin{eqnarray}\label{eq:b_equilibrium}
{\bm B}^{\rm eq}({\bm k}) = \left[vk - \int_0^{k_{\rm max}}\frac{dk'}{2\pi} k' f_-(k') V_1(k,k')\right]{\hat {\bm k}}~.
\end{eqnarray}
The ultraviolet cut-off $k_{\rm max}$ is required~\cite{b0footnote} because the integral is logarithmically divergent at large $k'$.
We set $k_{\rm max}$ to a value $\sim 1/a$, where $a \sim 0.246~{\rm nm}$ is graphene's lattice constant, 
set by the energy scale over which the Dirac model is valid.   
As expected ${\bm B}^{\rm eq}({\bm k})$ is oriented along ${\hat {\bm k}}$ and  
$|{\bm B}^{\rm eq}({\bm k})| = vk + \Sigma(k)$ depends only on $k$. 
Here we have introduced the pseudospin-dependent part of the Hartree-Fock self-energy
\begin{equation}\label{eq:equilibrium_HF_self_energy}
\Sigma(k)=- \int_0^{k_{\rm max}}\frac{dk'}{2\pi} k' f_-(k') V_1(k,k')~.
\end{equation}
Note that the fact that quasiparticles in monolayer graphene have chirality $J=1$ has selected 
the first moment $V_1(k,k')$ of the Coulomb interaction.

For convenience we assume a n-doped graphene layer and take $T=0$ so that 
\begin{eqnarray}\label{eq:identity}
f_-(k) = \frac{1}{2} [\Theta(k_{\rm F}-k) -1] &=& -\frac{1}{2}\Theta(k-k_{\rm F})~ \nonumber \\
f_+(k) - 1/2  &=& \frac{1}{2} \Theta(k_{\rm F}-k) 
\end{eqnarray}
where $k_{\rm F}$ is the conduction-band Fermi radius.  
The renormalized conduction $s = +$ and valence $s = -$ 
quasiparticle energies are therefore given by 
\begin{equation}\label{eq:equilibrium_HF_self_energy_numerics}
E_{s}(k) = s v k  - \frac{1}{4\pi}\int_{0}^{k_{\rm F}}dk' k' V_0(k,k') + s \Sigma(k) .
\end{equation}
with
\begin{equation} 
\label{eq:splititng}
 \Sigma(k) =  \frac{1}{4\pi}\int_{k_{\rm F}}^{k_{\rm max}}dk' k' V_1(k,k').
\end{equation}
The first interaction correction in Eq.~(\ref{eq:equilibrium_HF_self_energy_numerics}) is similar to the familiar self-energy correction which 
occurs in an ordinary two-dimensional electron gas.  When the Coulomb interaction is not screened it (famously and 
incorrectly) predicts a divergence of the quasiparticle velocity at the Fermi energy.  
Our focus here is on the second self-energy correction, exhibited explicitly in Eq.~(\ref{eq:splititng}),
which is responsible for enhanced splitting between valence and conduction bands.  
To account partially for screening and to avoid the well-known Fermi velocity artifact of mean-field-theory
in systems with long-range interactions
we have used a screened interaction potential of the form
\begin{equation}\label{eq:potential}
v_k = \frac{2\pi e^2}{\epsilon (\lambda q_{\rm TF} + k)}~,
\end{equation}
where $q_{\rm TF}=\alpha_{\rm ee}k_{\rm F}$ is the Thomas-Fermi screening vector and  
$\lambda$ is a control parameter which can be adjusted to represent the weaker screening at higher frequencies. 
($\alpha_{\rm ee} = e^2/\epsilon \hbar v$ is the graphene continuum-model fine structure constant.) 
We will mainly be interested in the contribution to the self-energy at $k \sim k_{\rm F}$ due to 
exchange interactions with states at $k' \gg k_{\rm F}$, for 
which $\lambda$ plays little role. 

\begin{figure}[t]
\begin{center}
\includegraphics[width=1.00\linewidth]{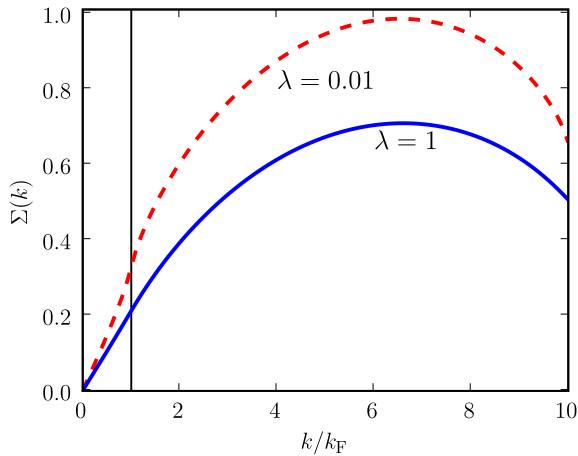}
\caption{(Color online) The pseudospin-dependent part of the Hartree-Fock self-energy $\Sigma(k)$ 
(in units of $\varepsilon_{\rm F}$) for monolayer graphene as a function of $k/k_{\rm F}$ (for wave vectors up to the cut-off) for $\alpha_{\rm ee}=0.5$ and $\Lambda=10$. The vertical (black) line indicates the point $k=k_{\rm F}$. The data labeled by the dashed (red) line correspond to $\lambda=10^{-2}$ in Eq.~(\ref{eq:potential}), while the data labeled by the solid (blue) line correspond to $\lambda=1$. Note the rapid change of $\Sigma(k)$ around $k=k_{\rm F}$ when $\lambda=10^{-2}$ which is an artifact of 
theories which do not accout for screening.  
The decrease in $\Sigma$ for $k$ near $k_{\rm max}$ is an artifact of the ultraviolet cut-off, but 
plays little role in the region of interest $k \sim k_{\rm F}$.\label{fig:one}}
\end{center}
\end{figure}

The dependence of $\Sigma(k)$ on $k/k_{\rm F}$ for $\alpha_{\rm ee}=0.5$ and $\Lambda=k_{\rm max}/k_{\rm F}=10$ is illustrated in Fig.~\ref{fig:one}. 
In this plot we have shown $\Sigma(k)$ for two different values of the control parameter $\lambda$ in Eq.~(\ref{eq:potential}): 
$\lambda=10^{-2}$, corresponding to an essentially unscreened Coulomb potential, and $\lambda=1$ 
corresponding to the Thomas-Fermi screened potential.  We see that the unphysical $\lambda=0$ anomaly at $k=k_{\rm F}$
emerges only at very small values of $\lambda$.  For non-zero values of $\lambda$, $\Sigma(k)$ is a smooth function
of $k$ and its contribution to the quasiparticle velocity at $k=k_{\rm F}$ can be obtained by expanding 
$\Sigma(k)$ for $k \ll k_{\rm max}$.
For $\lambda=0$ (unscreened Coulomb interactions) this expansion~\cite{expansionfootnote} can be performed analytically.  We find that
for $x > x'$ 
\begin{equation}\label{expansion1}
V_m(xk_{\rm F},x'k_{\rm F}) = \frac{2\pi e^2}{\epsilon k_{\rm F}}~{\bar V}_m(x,x')
\end{equation} 
where 
\begin{eqnarray} \label{expansion}
{\bar V}_m(x,x') &=& \int_0^{\infty}dt~J_{m}(tx)J_{m}(tx') \nonumber \\
&=&\frac{x'^{m}}{x^{m+1}}\frac{\Gamma(m+1/2)}{\Gamma(m+1)\Gamma(1/2)} \nonumber \\
&\times& ~_2F_1(m+1/2,1/2,m+1,x'^2/x^2).
\end{eqnarray}
Here $J_{m}(x)$ is a Bessel function of order $m$, 
$\Gamma(z)$ is the Gamma function, and $_2F_1(a,b,c,z)$ is the Gauss hypergeometric function. 
(The $x<x'$ result is obtained by interchange $x \leftrightarrow x'$.)
For $k_{\rm F} \ll k_{\rm max}$, a requirement for the utility of the Dirac continuum model 
of graphene, the leading term in the small $x'$ expansion of $V_m$ goes like $x'^m$.  As we explain below, this $m$-dependence 
is responsible for the main qualitative 
difference between the physics of quasiparticle mass renormalization 
in single-layer and bilayer graphene.

\begin{figure}[t]
\begin{center}
\includegraphics[width=1.00\linewidth]{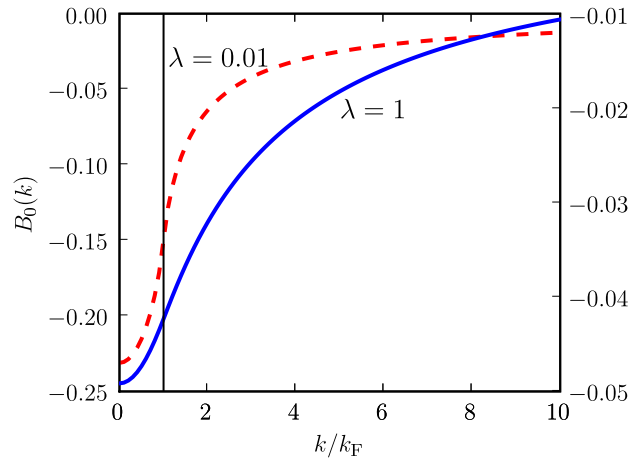}
\includegraphics[width=1.00\linewidth]{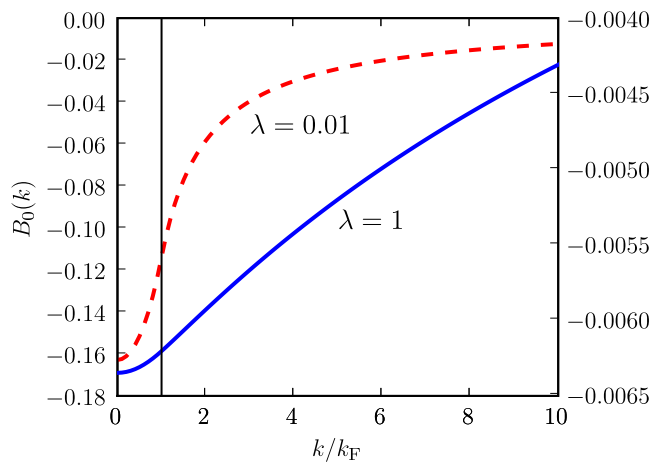}
\caption{(Color online) Pseudospin independent exchange energy contribution $B_0(k)$ 
(in units of $\varepsilon_{\rm F}$) as calculated from Eq.~(\ref{eq:bnot}) as a function of $k/k_{\rm F}$ for $\alpha_{\rm ee}=0.5$. 
The data labeled by the dashed (red) line correspond to $\lambda=10^{-2}$ in Eq.~(\ref{eq:potential}), while the data labeled by the solid (blue) line 
correspond to $\lambda=1$. Top panel: $\Lambda=10$. Bottom panel: $\Lambda=100$. 
In both panels the vertical (black) line indicates the point $k=k_{\rm F}$ and the right axis scale refers to the $\lambda=1$ data.\label{fig:two}}
\end{center}
\end{figure}

For $\Lambda = k_{\rm max}/k_{\rm F}$ large and $\lambda$ not too small, the 
quasiparticle mass renormalization is due entirely to the pseudospin dependent 
part of the self-energy.   (The pseudospin-independent contribution,
illustrated in Fig.~\ref{fig:two}, varies as $k^2$ at small $k$.) 
Neglecting $B_0$ we find that 
the renormalized velocity at $k=k_{\rm F}$ is 
\begin{eqnarray}\label{eq:vstar_over_v}
\frac{v^\star}{v} &=& 1 + \frac{1}{v}\left.\frac{\partial \Sigma(k)}{\partial k}\right|_{k=k_{\rm F}} \nonumber\\
&=& 1 + \frac{1}{2}\alpha_{\rm ee}\left.\frac{\partial f_\Lambda(x)}{\partial x}\right|_{x=1}~.
\end{eqnarray}
where 
\begin{eqnarray}\label{eq:fxLambda}
f_\Lambda(x) &=& \int_{1}^{\Lambda}dx' x' {\bar V}_1(x,x') \nonumber \\
&=& \frac{1}{2x^2}\int_{1}^{x}dx' x'^2~_2F_1(3/2,1/2,2,x'^2/x^2)\nonumber \\
&+&\frac{x}{2}\int_{x}^{\Lambda}dx' \frac{1}{x'}~_2F_1(3/2,1/2,2,x^2/x'^2)~.\nonumber \\
\end{eqnarray}
It follows that 
\begin{eqnarray}\label{eq:clean_derivative}
\left.\frac{\partial f_\Lambda(x)}{\partial x}\right|_{x=1} &=& 
\frac{1}{2}\int_{1}^{\Lambda}dx' \frac{1}{x'}~_2F_1(3/2,1/2,2,1/x'^2)\nonumber\\
&+&\frac{3}{8}\int_{1}^{\Lambda}dx' \frac{1}{x'^3}~_2F_1(5/2,3/2,3,1/x'^2)~.\nonumber\\
\end{eqnarray}
Because 
\begin{eqnarray}
&&\lim_{x'\to \infty}~_2F_1(3/2,1/2,2,1/x'^2) 
\nonumber\\
&=&\lim_{x'\to \infty}~_2F_1(5/2,3/2,3,1/x'^2) =1~,
\end{eqnarray}
we immediately see how the first term in Eq.~(\ref{eq:clean_derivative}) diverges logarithmically for large $\Lambda$: 
we find, to leading order in $\Lambda$ for $\Lambda \to \infty$
\begin{equation}
\lim_{\Lambda \to \infty} \left.\frac{\partial f_\Lambda(x)}{\partial x}\right|_{x=1} = \frac{1}{2}\ln(\Lambda)~,
\end{equation}
and thus we recover the well-known~\cite{velocityrenorm} asymptotic result for the velocity enhancement
\begin{eqnarray}
\lim_{\Lambda \to \infty} \frac{v^\star}{v} = 1 + \frac{\alpha_{\rm ee}}{4}\ln(\Lambda)~.
\end{eqnarray}

\section{Velocity Renormalization in Bilayer Graphene} 

Bilayer graphene has four atoms per unit cell and its $\pi$-electrons are therefore described by a model with four-component spinors.
In order to focus on the similiarities and differences between single-layer and bilayer velocity renormalization we use a 
commonly adopted two-component model for a bilayer which applies~\cite{mccann} at energies below the inter-layer tunneling scale 
and explains the bilayer's unusual~\cite{bilayerqhe} quantum Hall effect.  The two-component model for bilayer graphene has also 
been used to calculate the bilayer compressibility~\cite{viola_prl_2008} and the static noninteracting 
density-density linear-response function~\cite{hwang_prl_2008}.

The key differences between the two-band models
of single layer and bilayer graphene are: i) the band dispersion in the bilayer case is quadratic with an effective mass 
$m = t /(2 v^2)$ (where $t$ is the inter-layer tunneling and $v$ is the Fermi velocity in an isolated monolayer),
ii) the chirality is $J=2$ rather than $J=1$ for bilayer graphene (see below), and iii) the intra-layer [$V^{({\rm S})}_k = v_k$] and inter-layer 
[$V^{({\rm D})}_k = V^{({\rm S})}_k \exp(-k d)$] Coulomb interactions are different in the bilayer case.
In our discussion of bilayer graphene we use a Thomas-Fermi-like potential for $v_k$ as in Eq.~(\ref{eq:potential}), to avoid 
the well-known mean-field-theory artifacts in Coulombic systems.
We also employ a cut-off $k_{\rm max}$ for the bilayer case, although we shall see that its role is less essential.
(The two-band continuum model ultraviolet cutoff scale in bilayer graphene is set by $t$ which is 
smaller than the $\pi$-band width scale $\sim v/a$ appropriate in the single-layer case.)  
Our calculations for bilayers have screening ($\lambda$) and cut-off ($k_{\rm max}/k_{\rm F}\equiv \Lambda$) parameters as in the 
single layer case, and are in addition  
dependent on the dimensionless inter-layer distance parameter $\bar{d} = d k_{\rm max}$ which has a value~\cite{min_prb_2008} $\approx 0.2$.
The Thomas-Fermi screening vector for bilayer graphene is given by $q_{\rm TF} = m e^2 /\epsilon$ 
so that $q_{\rm TF}/k_{\rm F} = [t/(2v k_{\rm max})]\alpha_{\rm ee}\Lambda \equiv {\bar t} \alpha_{\rm ee} \Lambda $.  
Using $d= 3.35$~\AA~and $t=0.3~{\rm eV}$, 
we find that 
\begin{equation}\label{eq:bart}
\frac{q_{\rm TF}}{k_{\rm F}} = {\bar t} \alpha_{\rm ee} \Lambda  \simeq 0.38~\alpha_{\rm ee} \Lambda~.
\end{equation}

The mean-field theory calculations for the two-band model of bilayer graphene follow precisely
the same lines as in the single-layer case.  Eqs.~(\ref{eq:bnot}) and~(\ref{eq:equilibrium_B}) become:
\begin{equation}\label{eq:bnot2}
B_0({\bm k}) = - \int \frac{d^2{\bm k}'}{(2\pi)^2} V^{({\rm S})}_{{\bm k}-{\bm k}'} f_+(k')
\end{equation}
and
\begin{eqnarray}\label{eq:equilibrium_B2}
{\bm B}^{\rm eq}({\bm k}) &=& \frac{{\bm k}^2}{2 m}{\bm u}_{2}({\bm k}) \nonumber \\
&-& \int \frac{d^2{\bm k}'}{(2\pi)^2} V^{({\rm D})}_{{\bm k}-{\bm k}'} f_-(k'){\bm u}_{2}({\bm k}')~.
\end{eqnarray}
In Eq.(\ref{eq:equilibrium_B2}) ${\bm u}_{J}({\bm k})= (\cos(J\varphi_{\bm k}), \sin(J\varphi_{\bm k}))$ specifies the ${\bm k}$-dependence of the 
direction in pseudospin-space of the band-structure contribution to the effective magnetic field.  
In the single-layer case the chirality $J$ has the value $J=1$ and ${\bm u}_1({\bm k})={\hat {\bm k}} $.
It follows that only the pseudospin-dependent part of the Hartree-Fock self-energy differs between single-layer and 
bilayer cases: 
\begin{equation}\label{eq:equilibrium_HF_self_energy_bilayer}
\Sigma(k)=- \int_0^{k_{\rm max}}\frac{dk'}{2\pi} k' f_-(k') V^{({\rm D})}_2(k,k')~.
\end{equation}
Because the two low-energy sites which comprise the bilayer's pseudospin are in 
different layers, the pseudospin-dependent self-energy is proportional to the 
interlayer interaction $V^{({\rm D})}$.  The change from the $m=1$ angular component of the interaction in the single-layer case to the 
$m=2$ component in the bilayer case follows immediately from the $J=1$ to $J=2$ band-chirality change.
In Eq.~(\ref{eq:equilibrium_HF_self_energy_bilayer})
\begin{eqnarray}\label{eq:coeff_bilayer}
V^{({\rm D})}_m(k,k') &=& \int_0^{2\pi} \frac{d\theta}{2\pi} e^{-i m \theta}v(\sqrt{k^2+k'^2-2kk'\cos(\theta)}) \nonumber \\
&\times &\exp(-d \sqrt{k^2+k'^2-2kk'\cos(\theta)})~.
\end{eqnarray}
Using Eq.~(\ref{eq:identity}) we can rewrite Eq.~(\ref{eq:equilibrium_HF_self_energy_bilayer}) as
\begin{equation}
\Sigma(k) = \frac{1}{4\pi} \int_{k_{\rm F}}^{k_{\rm max}} dk' k' V^{({\rm D})}_2(k,k')~.
\end{equation}
In dimensionless units we have (scaling energies with $v k_{\rm F}$, which is {\it not} the Fermi energy!)
\begin{equation}\label{eq:equilibrium_HF_self_energy_numerics_dimensionless_bilayer}
\frac{\Sigma(x)}{v k_{\rm F}}= \frac{1}{2} \alpha_{\rm ee}
\int_{1}^{\Lambda}dx' x' {\bar V}^{({\rm D})}_2(x,x')~,
\end{equation}
where, as in the single-layer case, all wavevectors have been rescaled with $k_{\rm F}$, {\it i.e.} $x= k/k_{\rm F}$, $x' = k'/k_{\rm F}$, 
and where we have introduced the {\it dimensionless} interaction ${\bar V}^{({\rm D})}_m$, 
which is $V^{({\rm D})}_m$ measured in units of $2\pi e^2/(\epsilon k_{\rm F})$.

\begin{figure}[t]
\begin{center}
\includegraphics[width=1.00\linewidth]{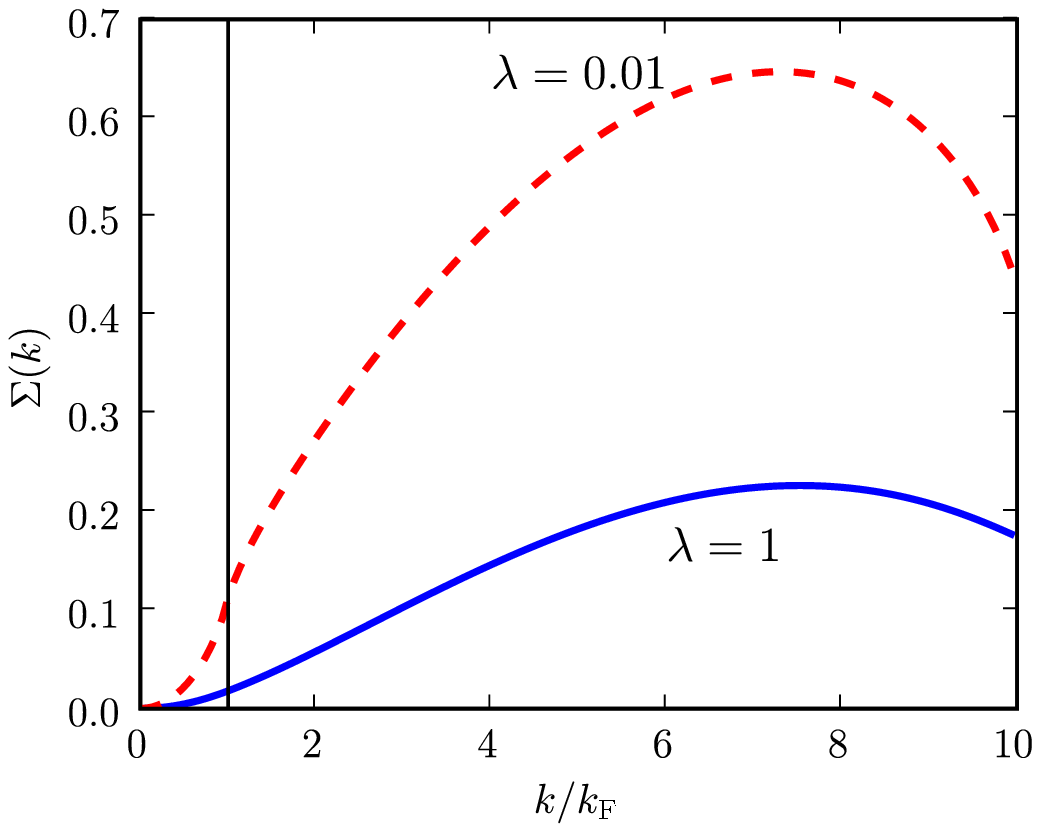}
\includegraphics[width=1.00\linewidth]{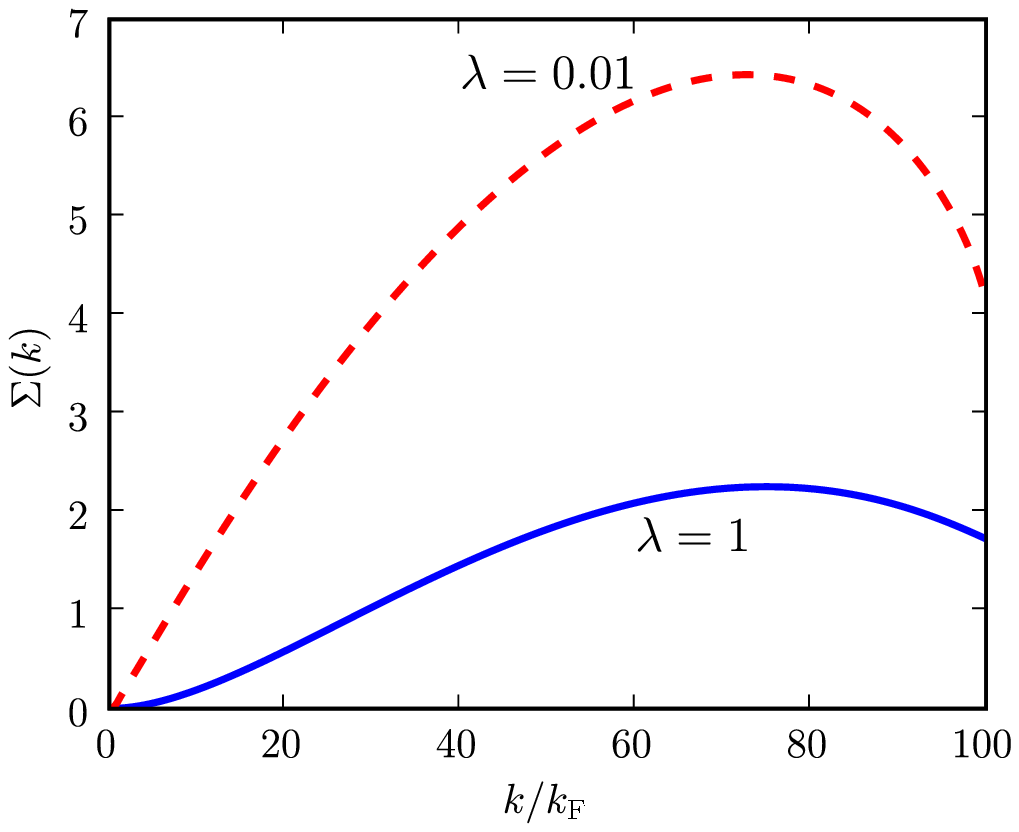}
\caption{(Color online) The pseudospin-dependent part of the Hartree-Fock self-energy $\Sigma(k)$ 
(in units of $ v k_{\rm F}$) for bilayer graphene as a function of $k/k_{\rm F}$ (for wavevectors up to the cut-off) for $\alpha_{\rm ee}=0.5$. 
The data labeled by the dashed (red) line correspond to $\lambda=10^{-2}$ in Eq.~(\ref{eq:potential}), while the data labeled by the solid (blue) line 
correspond to $\lambda=1$. Top panel:  $\Lambda=10$. The vertical (black) line indicates the point $k=k_{\rm F}$. Bottom panel: $\Lambda=100$.\label{fig:three}}
\end{center}
\end{figure}

The dependence of $\Sigma(k)$ on $k/k_{\rm F}$ for the bilayer case 
is illustrated in Fig.~\ref{fig:three} for $\alpha_{\rm ee}=0.5$ and $\Lambda=10$ and $100$.
In this plot we have shown $\Sigma(k)$ for two different values of the screening parameter $\lambda$ defined 
in Eq.~(\ref{eq:potential}): $\lambda=10^{-2}$, corresponding to an essentially unscreened Coulomb potential, and $\lambda=1$ corresponding to the Thomas-Fermi screened potential.
The small $k$ behavior is not strongly influenced by screening and can 
be understood analytically.
For large $\Lambda$ we can expand the exponential that enters in Eq.~(\ref{eq:coeff_bilayer}) 
in powers of ${\bar d}/\Lambda$:
\begin{eqnarray}
&&\exp\left( - \frac{\bar d}{\Lambda} \sqrt{x^2+x'^2-2xx'\cos(\theta)}\right) \nonumber \\
&\to& 1 - \frac{{\bar d}}{\Lambda} \sqrt{x^2+x'^2-2xx'\cos(\theta)} \nonumber \\
&+& \frac{1}{2}\left(\frac{{\bar d}}{\Lambda}\right)^2[x^2+x'^2-2xx'\cos(\theta)] +{\cal O}(({\bar d}/\Lambda)^3)~.\nonumber\\
\end{eqnarray}
For unscreened Coulomb interactions
\begin{eqnarray}\label{eq:coeff_bilayer_unscreened}
{\bar V}^{({\rm D})}_m(x,x') &=& \int_0^{2\pi} \frac{d\theta}{2\pi} e^{-i m \theta} \frac{1}{\sqrt{x^2+x'^2-2xx'\cos(\theta)}}\nonumber \\
&\times &\exp\left(- \frac{\bar d}{\Lambda} \sqrt{x^2+x'^2-2xx'\cos(\theta)}\right)~.\nonumber\\
\end{eqnarray}
This implies that the expansion of ${\bar V}^{({\rm D})}_m(x,x')$ starts at second order in ${\bar d}/\Lambda$ for $m \geq 1$: the term proportional to $\sqrt{x^2+x'^2-2xx'\cos(\theta)}$, which is first order in ${\bar d}/\Lambda$, is canceled by the denominator in the integrand in Eq.~(\ref{eq:coeff_bilayer_unscreened}). Thus the angular average in Eq.~(\ref{eq:coeff_bilayer_unscreened}) of this term 
would give us a finite result only for $m=0$. 

Disgarding this term, we obtain the following expansion of ${\bar V}^{({\rm D})}_m(x,x')$ in powers of ${\bar d}/\Lambda$ valid for $m \geq 1 $ :
\begin{eqnarray}\label{eq:crucial_expansion}
{\bar V}^{({\rm D})}_m(x,x') & \to & \left[1+ \frac{1}{2}\left(\frac{{\bar d}}{\Lambda}\right)^2(x^2+x'^2)\right]{\bar V}_m(x,x')  \nonumber \\
&-& \frac{1}{2} xx'  \left(\frac{{\bar d}}{\Lambda}\right)^2 \left[{\bar V}_{m+1}(x,x')+{\bar V}_{m-1}(x,x')\right]~.\nonumber\\
\end{eqnarray}
Here the coefficients ${\bar V}_m(x,x')$ are given by Eq.~(\ref{expansion}). Using this expression in Eq.~(\ref{eq:equilibrium_HF_self_energy_numerics_dimensionless_bilayer}) we find
\begin{widetext}
\begin{eqnarray}\label{eq:compact_expression}
\frac{\Sigma(x)}{v k_{\rm F}} &=& \frac{1}{2} \alpha_{\rm ee}
\int_{1}^{\Lambda}dx' x' {\bar V}_2(x,x')
+\frac{1}{4} \alpha_{\rm ee}\left(\frac{{\bar d}}{\Lambda}\right)^2 x^2 \int_{1}^{\Lambda}dx' x' {\bar V}_2(x,x')\nonumber\\
&+&\frac{1}{4} \alpha_{\rm ee}\left(\frac{{\bar d}}{\Lambda}\right)^2 \int_{1}^{\Lambda}dx' x'^3 {\bar V}_2(x,x')
-\frac{1}{4} \alpha_{\rm ee}\left(\frac{{\bar d}}{\Lambda}\right)^2 x \int_{1}^{\Lambda}dx' x'^2 [{\bar V}_1(x,x')+{\bar V}_3(x,x')]~.
\end{eqnarray}
\end{widetext}
To understand the limit $x \to 0$ of the bilayer Hartree-Fock self-energy in Eq.~(\ref{eq:compact_expression}) we use that
\begin{equation}\label{eq:leading_hypergeom}
{\bar V}_m(x \to 0,x') =  \frac{x^{m}}{x'^{m+1}} \frac{1 \cdot 3 \cdot 5 \cdot \dots \cdot (2m-1)}{2^m m!}.
\end{equation}
Inserting this expression in Eq.~(\ref{eq:compact_expression}), carrying out the integrations over $x'$, and neglecting terms that go to zero faster than $x^2$ for $x\to 0$, we finally find
\begin{eqnarray}\label{eq:compact_expression_final}
\frac{\Sigma(x \to 0)}{v k_{\rm F}} &=& \frac{3}{16}\alpha_{\rm ee} x^2~\frac{\Lambda-1}{\Lambda}\nonumber\\
&-& \frac{1}{32} \alpha_{\rm ee} x^2 {\bar d}^{\,2}~\frac{\Lambda-1}{\Lambda^2} + {\cal O}(x^3)~.
\end{eqnarray}
Note that the first term is finite in the limit $\Lambda \to \infty$ (zero doping limit), while the second term which contains the layer 
separation dependence goes to zero.  As in the single layer case, the dominant contribution to the exchange self-energy at small
wavevectors originates from interactions with states deep in the negative energy sea which are not sensitive to screening.

\begin{figure}[t]
\begin{center}
\includegraphics[width=1.00\linewidth]{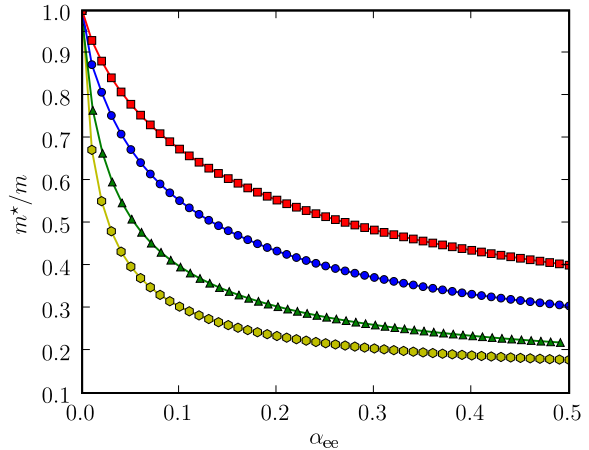}
\caption{(Color online) The Hartree-Fock renormalized mass $m^\star/m$ for bilayer graphene [calculated from Eq.~(\ref{eq:effective_mass})] 
as a function of $\alpha_{\rm ee}$ for $\lambda=10^{-2}$ and various values of $\Lambda$. From top to bottom, $\Lambda=10$ (squares), $20$ (circles), $50$ (triangles), and $100$ (hexagons). 
Note how interactions suppress the quasiparticle effective mass.\label{fig:four}}
\end{center}
\end{figure}

\begin{figure}[t]
\begin{center}
\includegraphics[width=1.00\linewidth]{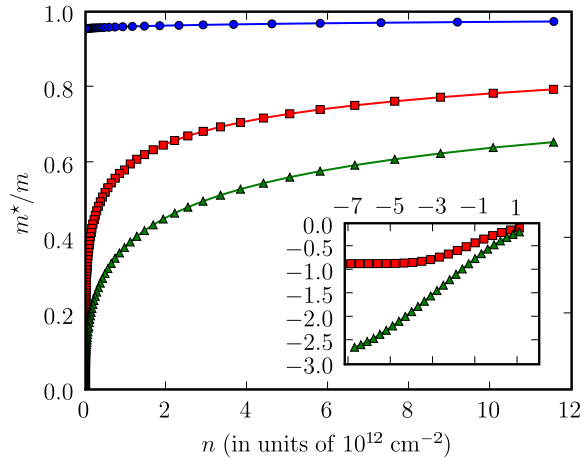}
\caption{(Color online) The Hartree-Fock renormalized mass $m^\star/m$ for bilayer graphene [calculated from Eq.~(\ref{eq:effective_mass})] 
as a function of density $n$ (in units of $10^{12}~{\rm cm}^{-2}$) for $\alpha_{\rm ee}=0.5$. The (blue) circles correspond to $\lambda=1$, the (red) squares to $\lambda=10^{-2}$, while the (green) triangles to $\lambda=10^{-4}$. 
In the inset we have reported $\log_{10}(m^\star/m)$ as a function of $\log_{10}(n)$ for $\lambda=10^{-2}$ and $\lambda=10^{-4}$.
Note that the bilayer renormalized mass always saturates at a finite value for $n \to 0$.\label{fig:five}}
\end{center}
\end{figure}

The renormalized mass $m^\star$ in bilayer graphene is defined {\it via}
\begin{eqnarray}
\frac{k_{\rm F}}{m^\star} & \equiv & \left.\frac{\partial}{\partial k}\left[\frac{{\bm k}^2}{2m}+B_0(k)+\Sigma(k)\right]\right|_{k=k_{\rm F}} \nonumber \\
&=& \frac{k_{\rm F}}{m} +\frac{1}{2} v \alpha_{\rm ee} \left.\frac{\partial \Gamma_\Lambda(x)}{\partial x}\right|_{x=1}~.
\end{eqnarray}
Restoring the pseudospin-independent self-energy contribution, which is identical in single layer and bilayer cases, 
we obtain that 
\begin{eqnarray}
\Gamma_\Lambda(x) &\equiv& \int_{1}^{\Lambda}dx' x' {\bar V}^{({\rm D})}_2(x,x')- 2 \int_0^{1}dx'~x' {\bar V}_0(x,x') \nonumber\\
&-& \int_{1}^{\Lambda}dx'~x' {\bar V}_0(x,x')~.
\end{eqnarray}
It follows that 
\begin{equation}\label{eq:effective_mass}
\frac{m^\star}{m} =\frac{1}{\displaystyle 1+\frac{1}{2}\alpha_{\rm ee}~{\bar t}~\Lambda~\left.\frac{\partial \Gamma_\Lambda(x)}{\partial x}\right|_{x=1}}~.
\end{equation}
where ${\bar t}$, defined above Eq.~(\ref{eq:bart}), is the same quantity that enters the Thomas-Fermi screening vector. 
We plot the ratio $m^\star/m$ obtained from this expression as a function of $\alpha_{\rm ee}$ for $\lambda=10^{-2}$ in Fig.~\ref{fig:four}
while in Fig.~\ref{fig:five} we plot $m^\star/m$ as a function of density for various values of the screening parameter 
$\lambda$.  In this last plot we have taken into account the spin and valley degeneracy factor $g=4$ of graphene by letting 
$q_{\rm TF} \to g q_{\rm TF}$.

\section{Summary and discussion} 

In summary, we have presented a comparative study 
of Fermi surface effective mass renormalization in single-layer and AB-stacked bilayer graphene.  
Interband exchange adds a new twist to electron-electron-interaction velocity renormalization physics compared to ordinary two-dimensional
electron systems~\cite{Giuliani_and_Vignale,asgari_prb_2005}, one which always leads to strongly {\it enhanced} velocities
and correspondingly {\it suppressed} masses. 
We have shown analytically [see Eq.~(\ref{eq:compact_expression_final})] and numerically (see Fig.~\ref{fig:five}, especially the inset) 
that although this mechanism causes the mass to approach zero at low carrier densities ($\Lambda \to \infty$) 
in single-layer graphene, the renormalized bilayer mass remains finite.  
The two cases are different because of the difference in the pseudospin-chirality of their
quasiparticles, $J=1$ in the single layer case and $J=2$ in the bilayer case.
Although the bilayer mass is finite, it is still estimated to be very strongly suppressed, depending on the approximation used for screening.

This behavior should be contrasted with what happens in the case of conventional parabolic-band electrons liquids in Si-MOSFETs~\cite{pudalov_prl_2002,shashkin_prb_2002} or in GaAs~\cite{tan_prl_2005}, where the quasiparticle velocity is generally suppressed by 
interactions (except in a narrow interval of very high densities -- small values of the Wigner-Seitz density parameter $r_s$ -- which depends in a complicated way on extrinsic effects such as the quantum-well finite thickness~\cite{asgari_prb_2005}).
The true value for $m^\star$ as a function of density for graphene bilayers probably occurs for a value of
the screening parameter $\lambda \sim 0.5$, which represents a 
compromise between using Thomas-Fermi theory and complete neglecting screening effects.  It should 
be emphasized however that a reliable numerical estimate for $m^\star$ 
as a function of density in bilayers lies outside the scope of this 
work and, as in the ordinary two-dimensional electron gas~\cite{asgari_prb_2005,asgari_prb_2006,markus_cond_mat_2008}, 
may ultimately require elaborate 
calculations that are informed by quantum Monte Carlo numerical studies. 

We conclude by commenting briefly on the possibility of broken symmetry states in single-layer and 
bilayer graphene systems. Enhanced quasiparticle velocities tend to increase the kinetic energy cost of moving the chemical potential
further away from the Dirac point.  The same physics which leads to velocity enhancement therefore
also tends to suppress spontaneous spin-polarization, the type of instability most often
contemplated~\cite{Giuliani_and_Vignale,nilsson_prb_2006} in ordinary low-carrier-density electron gas systems.  
Because of the pseudospin structure of single-layer and bilayer graphene, 
interactions instead favor chiral-symmetry-breaking-type scenarios, as we have explained elsewhere~\cite{min_prb_2008}.
Chiral symmetry breaking, which would lead to a spontaneous gap at the Dirac point and in 
the bilayer case to spontaneous charge transfer
between layers~\cite{min_prb_2008}, is most likely to occur in bilayer graphene.
If these instabilities occur, they are confined to a region of low carrier density~\cite{min_prb_2008} and the 
discussion of the current paper would apply only outside of this regime.  In bilayer graphene the occurrence or absence
of chiral symmetry breaking may depend on higher neighbor inter-layer hopping processes not included in the model discussed here. 
Recent lattice Monte Carlo calculations by Drut and L\"ahde~\cite{drut_prl_2009} suggest that these type of instabilities 
even occur for the $J=1$ single-layer case.  

\acknowledgements

M.P. was partly supported by the CNR-INFM ``Seed Projects".  Work in Austin was supported by the NRI SWAN project, 
by the Welch Foundation, and by the National Science Foundation under grant DMR-0606489.


\begin{thebibliography}{77}
\bibitem{expansionfootnote} 
	The artifacts associated with the neglect of screening do not appear in the small $k$ expansion.
\end{thebibliography}
\end{document}